# Probing quasi-long-range ordering by magnetostriction in monolayer $CoPS_3$

Qiye Liu[1,2,+], Le Wang[1,+], Ying Fu[1], Xi Zhang[5], Lianglong Huang[1], Huimin Su[1], Junhao Lin[1], Xiaobin Chen[6], Dapeng Yu[1], Xiaodong Cui[2], Jia-Wei Mei*[1,4], Jun-Feng Dai*[1,3]

1. Shenzhen Institute for Quantum Science and Engineering, and Department of Physics, Southern University of Science and Technology, Shenzhen, 518055, China
2. Department of Physics, The University of Hong Kong, Pokfulam 999077, Hong Kong
3. Shenzhen Key Laboratory of Quantum Science and Engineering, Shenzhen 518055, China
4. Shenzhen Key Laboratory of Advanced Quantum Functional Materials and Devices, Southern University of Science and Technology, Shenzhen 518055, China
5. Shannxi Institute of Flexible Electronics, Northwestern Polytechnical University, Xi'an 710072, China
6. School of Science and State Key Laboratory on Tunable Laser Technology and Ministry of Industry and Information Technology Key Lab of Micro-Nano Optoelectronic Information System, Harbin Institute of Technology, Shenzhen 518055, China

[+] The authors contribute to this work equally

* Corresponding authors:

daijf@sustech.edu.cn; meijw@sustech.edu.cn

**Abstract:**

**Mermin-Wagner-Coleman theorem predicts no long-range magnetic order at finite temperature in the two-dimensional (2D) isotropic systems, but a quasi-long-range order with a divergent correlation length at the Kosterlitz–Thouless (KT) transition for planar magnets. As a representative of two-dimensional planar antiferromagnets, single-layer $CoPS_3$ carries the promise of monolayer antiferromagnetic platforms for the ultimately thin spintronics. Here, with the aid of magnetostriction which is sensitive to the local magnetic order, we observe the signatured phonon mode splitting of below $T_{KT}$ in monolayer $CoPS_3$, revealing the presence of quasi-long-range ordering in XY-type antiferromagnet. Moreover, the ratio ($J'/J$) between the interlayer and intralayer interactions, which characterizes the 2D behaviors, is evaluated to be around 0.03 for the first time. Our results provide an efficient method to detect the quasi-long-range antiferromagnetic ordering in the two-dimensional magnets down to monolayer limit.**

As the scale of conventional ferromagnetic storage devices is approaching the limit of the critical domain size, growing interest has be concentrated on magnetic materials down to single atomic layer for high-density information storage and spintronic devices. It makes the 2D ferromagnetic materials a present hot spot along this approach[1-11]. Besides the ferromagnetic materials, antiferromagnetic materials play an unique role in spintronic devices thanks to the robustness of antiferromagnetic order to resist sizable external interference and zero net magnetization without any stray field[12]. Fundamentally both ferromagnetic and antiferromagnetic magnets are addressed in the similar frame of magnetic ordering. For Ising-type magnets where spin points in one direction either up or down[13], the phase transition are demonstrated experimentally in monolayer ferromagnet $Cr_2Ge_2Te_6$[1] and $CrI_3$[2] by magneto-optical Kerr effect (MOKE). The 2D XY-type system with spin confining within a-b plane is quite different from the Ising type as the Mermin-Wagner-Coleman theorem states that the long-range magnetic orders are suppressed at finite temperature[14,15]. Instead, the 2D XY-type magnetic materials could also be materialized in quasi-long-range magnetic ordered 2D systems. There is a presence of a phase transition at finite temperature to a phase without long-range order, but with a diverging in-plane correlation length. Such a phase transition is called the Kosterlitz–Thouless (KT) transition in 2D systems[16], which is corresponding to a gas of free vortices into binding of vortex-antivortex pairs at a temperature $T_{KT}$. This diverging correlation length could lead to a quasi-long range magnetic ordering and consequently a 2D magnet.

The emergence of transition metal phosphorus trisulfides ($MPS_3$, M = Mn, Fe, Co, Ni)[17-19], which is a class of potential 2D van der Waals (vdW) magnetic materials (Fig. 1a) provides a platform to explore the magnetic ordering down to monolayer limit. In contrast to $FePS_3$ and $(Mn/Ni)PS_3$ which are recognized as Ising-type and Heisenberg-type antiferromagnets in bulk form[20-24], respectively, $CoPS_3$ is one of 2D XXZ antiferromagnets[25]. $Mn^{2+}$ and $Ni^{2+}$ have the electronic configurations $3d^5$ and $3d^8$, respectively, in which the orbital degree of freedom for the d electrons is quenched. The spin-orbit coupling is negligible, and then $(Mn/Ni)PS_3$ are the almost isotropic Heisenberg-type antiferromagnets. $Fe^{2+}$ and $Co^{2+}$ have the electronic configurations $3d^6$ and $3d^7$, respectively, and the orbital degree is partially quenched. The spin-orbit coupling between spins and the unquenched orbitals accounts for the magnetic anisotropies in $FePS_3$ and $CoPS_3$, resulting in the Ising-type and XXZ-type antiferromagnets, respectively. The monolayer $CoPS_3$ as an exact 2D planar spin model likely exhibits the magnetic KT transition, however, experimental demonstration of the quasi-long-range order is challenging.

The spin-orbit coupling in $FePS_3$ and $CoPS_3$ due to the unquenched orbitals induces the structural change of magnetic materials under a magnetic phase transition, i.e., the magnetostrictive effect[26,27], which can be used as a predominant indicator of magnetic phase. For bulk $FePS_3$, the magnetostriction has been systematically studied by the low-temperature X-ray diffraction and neutron scatterings[28,29]. At the antiferromagnetic transition temperature in $FePS_3$, the lattice parameter c still linearly decreases continuously, whereas parameters a and b remarkably change in opposite way, where a-axes shrinks and b-axes extends, respectively (Fig.1b). This indicates that the magnetostriction of $CoPS_3$ can be used to probe the magnetic transition though not reported yet. Moreover, since it depends only on the local correlation, the magnetostriction should be sensitive to the quasi-long-range ordering even down to the monolayer limit. Along this line, in this work, we monitor the structural change by the Raman spectroscopy to probe the associated magnetic KT transition in the single- and few-layered $CoPS_3$. The structural change is characterized by the phonon mode splitting of Co-atom related vibration in Raman spectra, and the phase transition temperature $T_{KT}$ is observed to be ~100 K for the thin $CoPS_3$ from single layer to four layers, lower than the bulk antiferromagnetic transition temperature $T_N$=118 K. The associated magnetic transition in the few-layered $CoPS_3$ is also identified by the two-magnon signal. Our work demonstrates the implement of magnetostrictive effect in the exact 2D magnetic materials for the probe of the quasi-long-range order below the KT transition.

High quality $CoPS_3$ single crystals were grown by the chemical vapor transport (CVT) method (see method for details) and characterized with X-ray diffraction (Fig. S1) and X-ray photoemission (Fig. S2) techniques. The magnetic susceptibility ($\chi$) (Fig. S3a) and specific heat (Fig. S3b) measurements in $CoPS_3$ single crystal reveal a magnetic phase transition from paramagnetic to antiferromagnetic states at ~118 K as reported by A. R. Wildes et al[30]. Recent neutron diffraction[30] studies have revealed $CoPS_3$ has an intralayer antiferromagnetic structure, where ferromagnetic 'zig-zag' chains are formed along the a-axis, while adjacent chains are coupled anti-parallelly along the b-axis (Fig. 1b). As shown in Fig.1a, monolayer $CoPS_3$ has a hexagonal structure for moments on $Co^{2+}$ irons; for few layered $CoPS_3$, adjacent layers are weakly coupled to each other by vdW forces along the c axis. Therefore, $CoPS_3$ flakes can be mechanically exfoliated from single crystal onto $SiO_2$/Si substrates using sticky tape, which is a common method to produce transition metal dichalcogenides (TMDCs) sheets[31,32]. The optical and atomic force microscope (AFM) images of monolayer $CoPS_3$ is shown in Fig. 2a. The thickness of monolayer $CoPS_3$ is around 0.87 nm. In addition, the thickness of bilayer, trilayer, and quadlayer $CoPS_3$ is 1.49, 2.26, and 2.86 nm (Fig. S4), respectively.

The polarized Raman spectroscopy[33,34] has been proved to be sensitive to a structural change even for atomically thin samples. Hence, in this study, we apply the polarized Raman spectroscopy to characterize the magnetic state and crystal structural change in monolayer $CoPS_3$. More details about polarized Raman measurements can be found in method section. Fig. 2b shows the representative circularly polarized Raman spectra of a monolayer $CoPS_3$ at room temperature (T=295 K) and low temperature (T=25 K). Here we excite the samples with a left-handed (L) circularly polarized light, and monitor the left-handed (L) and right-handed (R) circularly polarized components of Raman signals.

Five unambiguous peaks are clearly identified at room temperature. Based on $D_{3d}$ point group, two modes at 155 ($P_2$) and 282 ($P_6$) cm$^{-1}$ are assigned to the double-degenerate $E_g$ modes, whereas three peaks at 242 ($P_5$), 384 ($P_7$) and 585 ($P_9$) cm$^{-1}$ are non-degenerate $A_g$ vibration modes. With the aid of density-functional theory (DFT) calculations (Table S1), the low-frequency peak ($P_2$) is attributed to in-plane Co-atom related vibrations. While $P_5$, $P_6$ and $P_7$ modes originate from out-of-plane and in-plane vibrations of S-S atoms and $P_9$ is the out-of-plane vibrations of P-P atoms. Other Raman peaks listed in the Table S1, including $P_1$, $P_4$, $P_8$, cannot be detected in monolayer $CoPS_3$ due to relatively weak signal strength. While for bulk $CoPS_3$, these three peaks centered at 111 ($P_1$), 229 ($P_4$), and 559 ($P_8$) cm$^{-1}$ can be clearly resolved (Fig. S5a), which are assigned as double-degenerate $E_g$ modes. Among them, $P_1$ is also related to in-plane Co-atom related vibrations. In addition, $P_3$ corresponds to the non-active $A_{2g}$ mode in $D_{3d}$ point group at room temperature. These findings are consistent with the results in literature[35], which also imply the high quality of our experimental samples.

At low temperature, a remarkable change can be found at the phonon peak $P_2$, where the double-degenerate $E_g$ mode at 155 ($P_2$) cm$^{-1}$ at 295 K splits into two components with peak positions at 149 ($P_2'$) and 157 ($P_2''$) cm$^{-1}$ at 25 K in the circularly cross-polarized (LR) component of Raman spectrum, respectively. The changes of the peak position of $P_2'$ and $P_2''$ show opposite trends with respective to the $P_2$ peak, i.e., redshift for $P_2'$ and blueshift for $P_2''$. Meanwhile there is no noticeable splitting in another $E_g$ mode (peak $P_6$) corresponding to the in-plane vibrations of S-S atoms. In addition, a new peak centered at 149 ($P_2'$) cm$^{-1}$ appear in the co-polarized (LL) component of Raman spectrum, which is absent in the corresponding spectrum at 295 K. The similar peak splitting can also be observed in the $P_1$ peak, which is also corresponding to Co-atom related vibration mode. For bilayer, trilayer, quadlayer and even bulk CoPS$_3$, the $P_2$ peak splitting is clearly observed as indicated by the dashed lines in Fig. 2c. The corresponding Raman signals ($P_2'$) in circularly co-polarized (LL) component gradually rise with the increasing number of layers. Moreover, for the $P_5$ an $P_7$ peaks in bulk CoPS$_3$ (Fig. S5), an extra component of LR configuration emerges in the Raman spectra at 25 K, while a new peak centered at 187 cm$^{-1}$ ($P_3$) also appears in LR configuration at 25 K.

According to Raman tensor analysis (method section), the structure of monolayer CoPS$_3$ is orthohexagonal at high temperature, and the Raman-active phonon modes can be described as the point group $D_{3d}$. At low temperature, the double degenerate $E_g$ mode tends to split into $A_g$ and $B_g$ modes. While the $A_{1g}$ mode turns to be an $A_g$, which responds differently from that emerged from the $E_g$ mode. The invisible $A_{2g}$ mode turns to be a $B_g$ mode. The symmetry analysis of the mode evolution from high temperature to low temperature implies that the Raman tensors shifts from the point group $D_{3d}$ to $C_{2h}$ due to the structural distortion in monolayer CoPS$_3$ system[33], as illustrated in Fig. 1b. Based on the Raman tensors of $C_{2h}$ point group (see method section), we except that for the $B_g$ mode, the left-handed circularly polarized excitation generates the right-handed circularly polarized Raman signal, or vice versa. Whereas the $A_g$ mode will appear in both components, i.e., the left-handed and right-handed components, under a left-handed (or right-handed) circularly polarized excitation. These are consistent with our experimental observation. Following this rule, the $P_2'$ ($P_1''$) is assigned as an $A_g$ phonon mode, while the $P_2''$ ($P_1'$) belongs to a $B_g$ phonon mode at low temperature. The layer-independent peak splitting (Fig. 2c) also indicates that intralayer structural distortion dominates the Raman response instead of the tiny interlayer translation.

To further confirm that the peak splitting or the structural distortion is related to magnetic behaviors in monolayer limit, we also inspect the linearly polarized Raman spectra as a function of temperature in the monolayer CoPS$_3$. In this case, the P$_2$' and P$_2$" will emerge at cross-polarization (XY) and co-polarization (XX) scattering configurations, respectively, so that we can exactly extract the peak position at different temperatures. As shown in Fig. 3a, we can clearly see that the split doubly degenerated P$_2$ peaks gradually merge into one peak with increasing temperature. The frequency difference ($\Delta P_2 = P_2^{''} - P_2^{'}$) between two split peaks (black squares in Fig. 3b) remains around 7.5 cm$^{-1}$ below ~ 65 K. After that, it dramatically decreases and approaches zero above ~100 K. In addition, Fig 3b also shows that the splitting of the P$_2$ for samples of 2-4 layers exhibits the similar temperature dependence as that of the monolayer one with the same transition temperature of ~100 K ($T_{KT}$). On the contrary, the structural transition temperature of few-layered CoPS$_3$ (1-4 layers) is lower than that of the bulk at 118 K. It suggests that these thin samples have already exhibited 2D behaviors. For the bulk CoPS$_3$, the peak splitting $\Delta P_2$ (red diamonds in Fig. 3b) tracks the magnetic scattering intensity in the neutron scattering measurements (short dashed line in Fig. 3b), indicting the magnetostriction in the bulk. However, the connection between the structural change and the magnetic transition for few-layered CoPS$_3$ is not established yet. It is our next task as demonstrated in two-magnon scattering in Fig. 4, which will be discussed later.

Further evidence for structural distortion due to magnetostriction can be found in the temperature dependence of the peak P$_5$ near 250 cm$^{-1}$ in monolayer CoPS$_3$. Fig. 3a shows that the peak energy starts to redshift as temperature decreases below ~86 K (dashed line in Fig. 3a), which is near the $T_{KT}$ for thin CoPS$_3$. For bulk samples, the peaks P$_5$ near 250 cm$^{-1}$ (green triangles in Fig. S5b) and P$_7$ near 386 cm$^{-1}$ (orange squares in Fig. S5b) exhibit the similar temperature-dependent trend as monolayer one, however, the temperature threshold of the peak energy redshifts and blueshifts is $T_N$ instead of $T_{KT}$, respectively. Based on the DFT calculations on the vibration modes of the P$_5$ and P$_7$, they originate from the out-of-plane vibration of S-S atoms without the participation of any magnetic atom. Therefore, this temperature-dependent Raman shift in monolayer and bulk CoPS$_3$ must be indirectly related to magnetostrictive effect. We suggest that the distortion of hexagonal structure formed by Co atoms occurs at low temperature changes the lattice parameters and influences the Raman mode of $(P_2S_6)^{4-}$ bipyramid structure.

Fig. 4a shows the temperature dependence of the Raman spectra in a bilayer $CoPS_3$ measured in the cross-polarization (XY) configuration. We can clearly observe that the two-magnon signals appear at the central wavenumber of 350 and 600 cm$^{-1}$ with an intriguing broad bandwidth at low temperature, which gradually disappears as temperature increases. These observations are the typical phenomena of antiferromagnetic materials, and the similar behaviors can be observed in $NiPS_3$[36]. Fig. 4c summarizes the normalized integrated intensities of two-magnon Raman signals as a function of temperature for samples with different thicknesses. For bulk samples, the temperature dependence of the two-magnon Raman is consistent with that of Bragg peak intensity by neutron diffraction measurements in ref [30]. As the thickness of samples decreases, the normalized integrated intensities obviously diverge from that of bulk one. In addition, as shown in Fig. 4a, the Fano resonance of $P_8$ peak can also be observed at low temperature in cross-polarization (XY) configuration of Raman spectra of the bilayer sample, which gradually rise as the number of layers increase. The 1/q (see SI for definition), which represents the strength of coupling between a discrete excitation and a two-magnon excitation, is extracted to be around 0.7 for the bulk material below $T_N$. As shown in Fig. S6, the coupling strength of Fano resonance dramatically increases in comparison with the values above $T_N$ due to the presence of antiferromagnetic orders. Therefore, we establish the connection between the structural change and the magnetic transition for few-layered $CoPS_3$, indicating that the peak splitting in Fig. 2 and Fig. 3 is an indicator of the magnetic transition.

For the few layers of $CoPS_3$, the system is described by the two-dimensional *XY* model and undergoes a finite temperature transition at $T_{KT}$. Due to the strong magnetostriction effect, the onset of the magnetic order is accompanied by the peak splitting of the phonon modes at 155 ($P_2$) cm$^{-1}$. We expect that the lattice distortion and hence the peak splitting is proportional to the average amplitude of the spin-pair $\sum_{\langle ij \rangle} \langle S_i \cdot S_j \rangle$. Although the staggered magnetization is zero, $\langle |S_i| \rangle = 0$, for the monolayer $CoPS_3$, the in-plane correlation length goes to an unusual exponential divergence as $T$ approaches $T_{KT}$[37], $\xi(T) \sim \exp(B(T/T_{KT} - 1)^{-1/2})$ with $B \sim \pi/2$. Such a divergent correlation length gives rise to the non-zero summation in the spin-pair $\sum_{\langle ij \rangle}\langle S_i \cdot S_j \rangle$, resulting the peak splitting of the phonon modes. We notice that the transition temperature remains unchanged from monolayer to 4-layer one, indicating the two-dimensional behavior in the few-layered samples.

For the bulk material, the small interlayer coupling of the system slightly modifies the transition temperature

$$T_N = T_{KT} + C/\left(ln(J'/J)\right)^2 T_{KT} \quad {}^{38}$$

where C~2 and J'/J is the ratio between the interlayer and intralayer interactions. As $T_{KT} = 100\ K$ for thin samples and $T_N = 118\ K$ for bulk in the measurements, we can estimate $J'/J$ =0.03. For the bulk case, a 2D XY-like behavior is expected in the temperature range between $T_{KT}$ and $T_N$, when the 3D fluctuation is weak. Below $T_{KT}$ (green region in Fig. 3b), the system is characterized as the quasi-long range magnetic orders, while it is the paramagnetic state above $T_N$ (purple region in Fig. 3b). Between them (orange region in Fig. 3b), the interlayer interaction (J') becomes important for bulk materials, indicating the 3D fluctuations in this regime. Moreover, the effect of the creation of the vortex-antivortex pairs leads a very rapid decrease of the intensity in the neutron scattering and two-magnon Raman scattering as observed in the experiments. The matrix element of the two-magnon Raman scattering is also proportional to $\sum_{\langle ij \rangle}\langle S_i \cdot S_j \rangle$, accounting for the same two-magnon intensity onset temperature as $T_{KT}$ instead of $T_N$.

In conclusion, we have successfully demonstrated paramagnetic-antiferromagnetic phase transition in monolayer $CoPS_3$ with the polarization-resolved Raman spectroscopy. The splitting of double-degenerate $E_g$ phonon mode of Co-atom related vibration and the appearance of new peaks ($P_2$') in LL configuration at low temperature clearly indicate the magnetostrictive effect in the presence of quasi-long-range ordering below $T_{KT}$ in a 2D XY-type antiferromagnet. Our results will be helpful in understanding the underlying mechanism of magnetostrictive effect in 2D materials and clearing some obstacles for manufacturing future antiferromagnetic storage devices.

**Methods**

**Sample growth:** $CoPS_3$ single crystals were successfully grown by the chemical vapor transport (CVT) method. Cobalt powder (99.998%), phosphorus flakes (99.999%) and sulfur powder (99.9995%) in 1 : 1 : 3 molar ratio were mixed together and put into a silica tube with a length of 150 mm and an inner diameter of 17 mm. After the tube was evacuated down to $10^{-3}$ Pa and sealed, it was placed in a two-zone horizontal tube furnace, where the source and growth zones were raised to 873 K and 823 K, respectively. After holding at the temperatures for 7 days, the hexagonal dark gray crystals with lateral dimensions up to several millimeters can be obtained. The optical image in the inset of Fig. S1 shows the crystal is characteristic of a hexagonal crystal structure with metallic grey color. Typical crystal size is around 2 x 3 mm$^2$.

**Magnetic susceptibility and specific heat:** The magnetic susceptibility ($\chi$) was performed on a Quantum Design (QD) MPMS3 SQUID magnetometer and the specific heat ($C_p$) was measured in a physical property measurement system (PPMS) using a thermal-relaxation method. The experimental results are shown in Fig. S3.

**Polarized Raman spectroscopy:** Polarized Raman spectra were performed using a homemade polarization-resolved Raman spectroscopy in the back-scattering geometry. A solid-state laser at 532 nm was employed to excite CoPS$_3$ samples. The laser beam passed through a laser line filter, a beam splitter, and was focused onto the CoPS$_3$ samples by a 50x objective. The back-scattered light was collected by the same objective, passed through the beam splitter and two 532 nm edge filters, and then was focused on the entrance slit of a spectrometer with a 1800 grooves mm$^{-1}$ diffraction grating and a nitrogen-cooled charge coupled devices (CCD). To conduct the polarization-dependent measurement, two polarizers were put after the laser line filter and before the spectrometer with parallel- and crossed-polarization configurations, respectively. For circularly polarized Raman measurement, a quarter waveplate was located before the objective to tune the polarized state of excited light from linear polarization to circular polarization. Two circular components of Raman signal passed through the same waveplate and changed to two orthogonal linear polarized states. By combining a ½ waveplate and a polarizer before the spectrometer, we can extract two circular Raman signals, respectively.

**Raman tensor:** At high temperatures, the structure of CoPS$_3$ is orthohexagonal, and the Raman-active phonon modes can be labeled by the point group D$_{3d}$

$$\Gamma_{Raman} = 3A_{1g} + 5E_g + 2A_{2g}$$

Note the A$_{2g}$ is not visible for the D$_{3d}$ group in the Raman scattering, however, it turns out to have the visible intensity for the C$_{2h}$ point group for the monoclinic structure. The Raman tensors for the hexagonal structure at high temperature are given as

$$\tau_{A_{1g}} = \begin{pmatrix} a & 0 & 0 \\ 0 & a & 0 \\ 0 & 0 & b \end{pmatrix}, \tau_{E_g}^{(1)} = \begin{pmatrix} c & 0 & 0 \\ 0 & -c & d \\ 0 & d & 0 \end{pmatrix}, \tau_{E_g}^{(2)} = \begin{pmatrix} 0 & -c & -d \\ -c & 0 & 0 \\ -d & 0 & 0 \end{pmatrix}, \tau_{A_{2g}} = \begin{pmatrix} 0 & e & 0 \\ -e & 0 & 0 \\ 0 & 0 & 0 \end{pmatrix}$$

which explains the angle independence of the linearly polarized Raman scattering behavior at high temperatures. At low temperature, if a structural distortion occurs, we expect that the Raman tensors will be expressed as the C$_{2h}$ point group

$$\tau_{A_g}^1 = \begin{pmatrix} a + a1 & 0 & 0 \\ 0 & a & 0 \\ 0 & 0 & b \end{pmatrix}, \tau_{A_g}^2 = \begin{pmatrix} c + c1 & 0 & 0 \\ 0 & -c & d \\ 0 & d & 0 \end{pmatrix}, \tau_{B_g}^1 = \begin{pmatrix} 0 & -c & -d \\ -c & 0 & 0 \\ -d & 0 & 0 \end{pmatrix}, \tau_{B_g}^2 = \begin{pmatrix} 0 & e & 0 \\ -e & 0 & 0 \\ 0 & 0 & 0 \end{pmatrix}$$

Under the excitation of a circularly polarized light, the Raman intensity for different A$_g$ and B$_g$ modes at C$_{2h}$ point group can be expressed as following:

$$\text{LL: } I_{A_g}^1 = \frac{(2a+a1)^2}{2}, I_{A_g}^2 = \frac{c1^2}{2}, I_{B_g}^1 = I_{B_g}^2 = 0 \qquad (1)$$

$$\text{LR: } I_{A_g}^1 = \frac{a1^2}{2}, I_{A_g}^2 = \frac{(2c+c1)^2}{2}, I_{B_g}^1 = \frac{(2c)^2}{2}, I_{B_g}^2 = \frac{(2e)^2}{2} \qquad (2)$$

where LL means the left-handed circularly polarized state of excitation and scattering Raman signal, LR means the opposite state between them, i.e., left circularly polarized state for excitation and right circularly polarized state for Raman signal. The superscripts 1 and 2 represent different Raman modes.

## Acknowledgements

Q.L. and L.W. contribute this work equally. J.D. and J.M. conceived the project. Q.L., H.S. and J.D. designed and performed the experiments. L.W. and L.H. provided the experimental samples. Y.F. and J.M. provided the theoretical supports. X.Z. and J. L. conducted the AFM measurements. All authors discussed the results and co-wrote the paper. The authors would like to thank Prof. A R Wildes from Institut Laue-Langevin and Prof. Haizhou Lu from SUSTech for helpful discussions. J.F. acknowledges the support from the National Natural Science Foundation of China (11974159). J.W.M was partially supported by the program for Guangdong Introducing Innovative and Entrepreneurial Teams (No. 2017ZT07C062), and Shenzhen Key Laboratory of Advanced Quantum Functional Materials and Devices (No. ZDSYS20190902092905285). J.L. acknowledges the support from National Natural Science Foundation of China (Grant No.11974156) and the Guangdong International Science Collaboration Project (Grant No. 2019A050510001).

## Conflict of Interest

The authors declare no conflict of interest.

## Keywords

**Figures and captions**

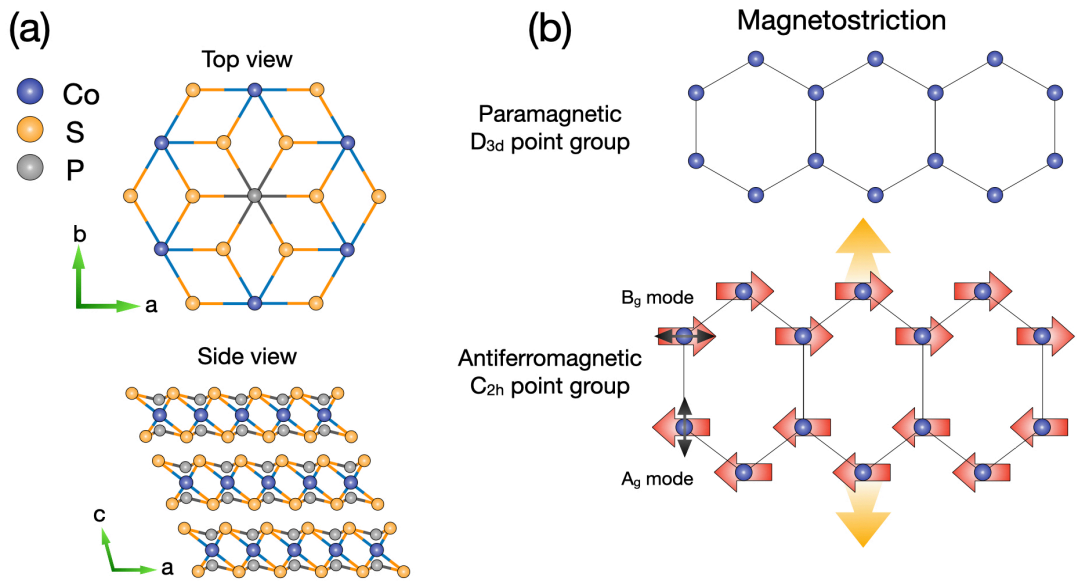

Figure 1: (a) Top view of monolayer $CoPS_3$ and side view of crystal structure of bulk $CoPS_3$. Blue, orange and gray balls represent Co, S and P atoms, respectively. (b) Magnetostriction in monolayer $CoPS_3$, where a structural distortion is induced at antiferromagnetic state. In this case, it belongs to $C_{2h}$ point group. At paramagnetic state, crystal structure is close to being orthohexagonal and is described by $D_{3d}$ point group. Red arrows represent the spin orientations of Co atoms at antiferromagnetic state and orange ones represent extension of lattice structure. $A_g$ and $B_g$ vibration modes are also indicated by the black arrows, which are parallel and perpendicular to spin orientation, respectively.

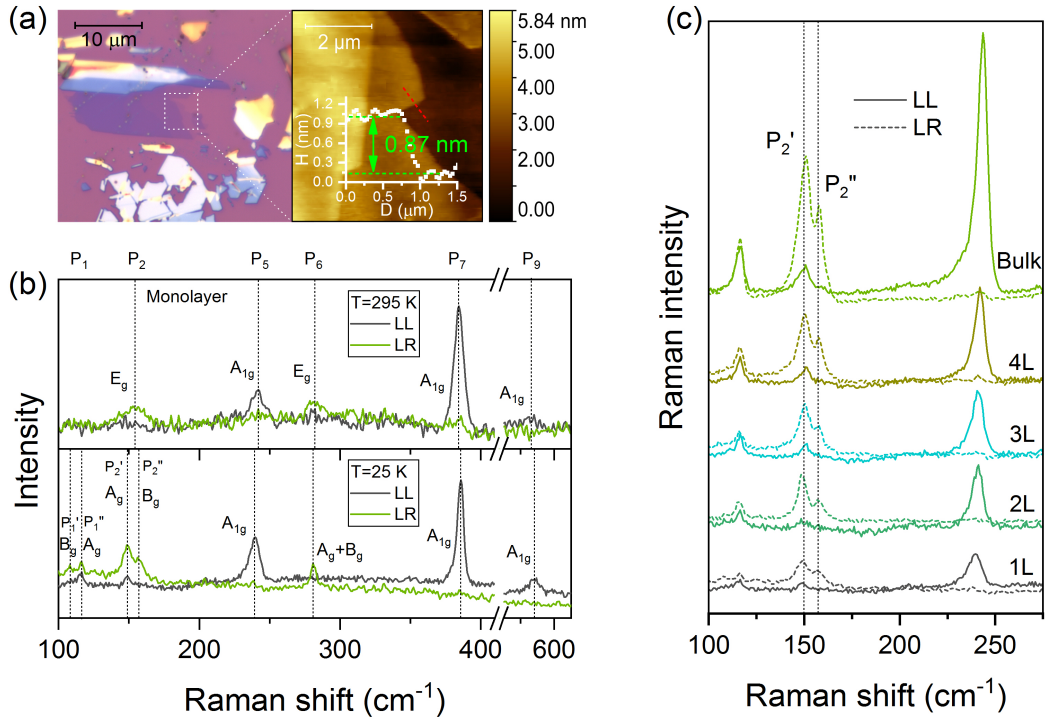

Figure 2: (a) Optical and AFM image of monolayer $CoPS_3$. The height of the monolayer is about 0.87 nm. (b) Circularly polarized Raman spectra of the monolayer $CoPS_3$ measured at 295 K and 25 K in circularly co-polarization (LL) and cross-polarization (LR) scattering configurations, respectively. LL (LR): the first letter represents the polarized state of excited light, whereas the second letter represents the polarized state of scattering Raman signal. And L and R means left-handed and right-handed circularly polarized light, respectively. The dashed lines indicate the position of peaks, which as labeled as $P_1$, $P_2$, $P_5$, $P_6$, $P_7$, and $P_9$. The corresponding Ramon modes are summarized in Table S1 and shown in the figure. Compared with the spectra at 295 K, $P_2$ peak splits into two components at low temperature, which are labeled as $P_2'$ and $P_2''$. (c) Raman spectra of $CoPS_3$ with different layer numbers in LL (dashed lines) and LR (solid lines) configurations at 25 K, respectively. The splitting of $P_2$ peak still can be observed in all the samples as indicated by two dashed lines.

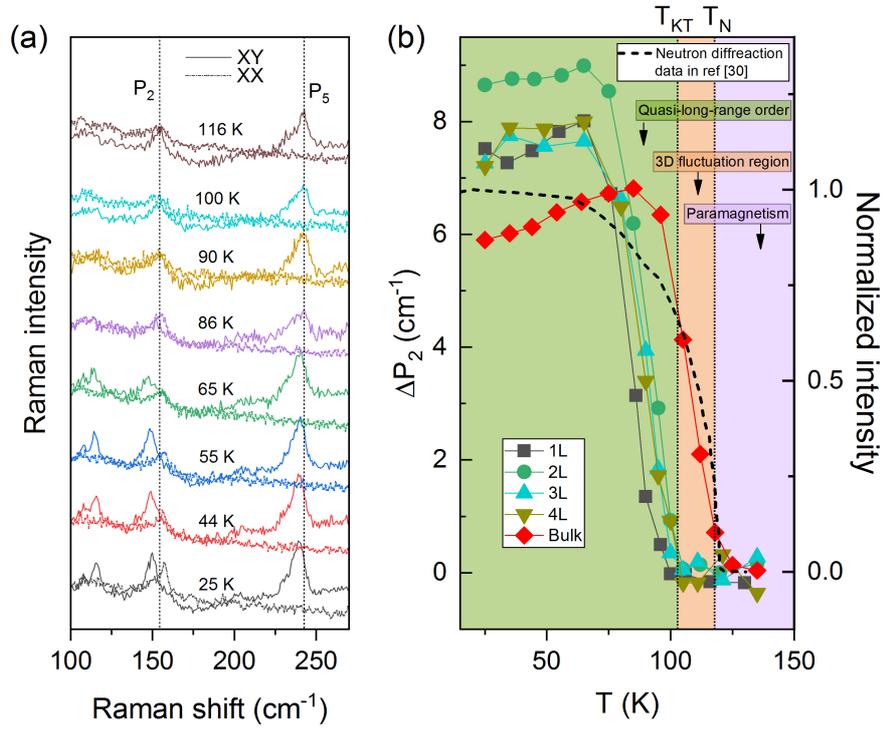

Figure 3: (a) Raman spectra of monolayer CoPS$_3$ at several representative temperatures in co-polarization (XX) (dashed lines) and cross-polarization (XY) (solid lines) configurations, respectively. The dashed lines indicate the peak position of P$_2$ and P$_5$ at room temperature. (b) ΔP$_2$ as a function of temperature in CoPS$_3$ with different layer numbers and neutron diffraction data in ref [30]. The dashed line indicates the position of T$_{KT}$ and T$_N$, respectively. Three regions are filled with different colors and labeled as quasi-long-range order, 3D fluctuation region and paramagnetism, respectively.

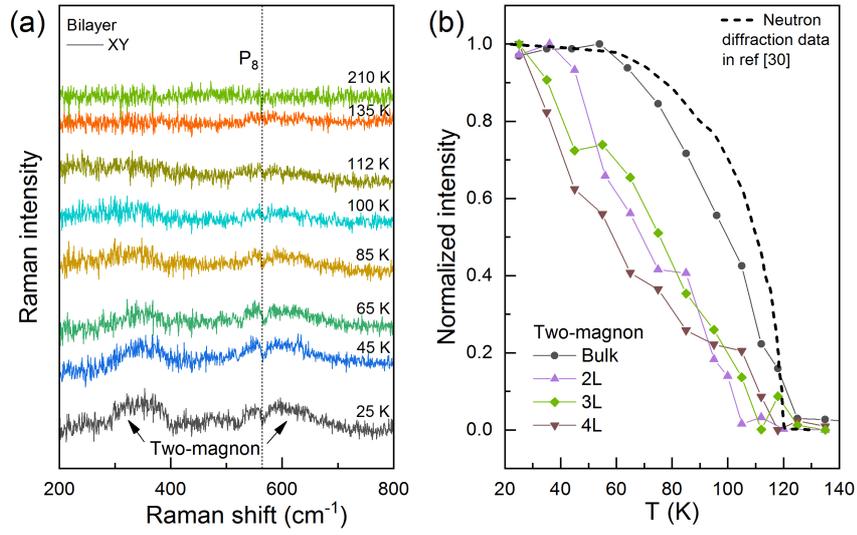

Figure 4: (a) Raman spectra of bilayer $CoPS_3$ in the range of 200-800 $cm^{-1}$ at several representative temperatures in cross-polarization (XY) configuration. Two-magnon peaks are indicated by the black arrows. Peak $P_8$ is also labeled in the figures. To clearly exhibit two-magnon peaks, the phonon peak $P_6$ around 250 $cm^{-1}$ is eliminated. (b) Normalized integrated intensity of two-magnon scattering as a function of temperature in $CoPS_3$ with different layer numbers and neutron diffraction data in ref [30] as a function of temperature.